\documentclass[prl,showpacs,twocolumn,floatfix,bibnotes]{revtex4}
\usepackage{here}
\usepackage{dcolumn}
\usepackage{graphicx}
\usepackage{bm}

\begin{document}

\title{Melting of Discrete Vortices via Quantum Fluctuations}

\author{Chaohong Lee, Tristram J. Alexander, and Yuri S. Kivshar}

\affiliation{Nonlinear Physics Centre and ARC Centre of Excellence
for Quantum-Atom Optics, Research School of Physical Sciences and
Engineering, Australian National University, Canberra ACT 0200,
Australia}


\date{\today}

\begin{abstract}
We consider nonlinear boson states with a nontrivial phase
structure in the three-site Bose-Hubbard ring, {\em quantum
discrete vortices} (or {\em q-vortices}), and study their
``melting" under the action of quantum fluctuations. We calculate
the spatial correlations in the ground states to show the
superfluid-insulator crossover and analyze the fidelity between
the exact and variational ground states to explore the validity of
the classical analysis. We examine the phase coherence and the
effect of quantum fluctuations on q-vortices and reveal that the
breakdown of these coherent structures through quantum
fluctuations accompanies the superfluid-insulator crossover.
\end{abstract}

\maketitle

Vortices are fundamental objects in physics which appear in
different fields including phase singularities in
optics~\cite{soskin} and circulating bosons in Bose-Einstein
condensates (BECs)~\cite{bec}. Periodic lattices such as periodic
photonic structures for light waves or optical lattices for BECs can
modify strongly the wave propagation, and may support novel types of
vortex states termed {\em discrete vortex
solitons}~\cite{discrete,lena}. Such discrete vortices describe {\em
spatially localized circular energy flows}. They have been studied
theoretically and observed experimentally in optically-induced
photonic lattices as stable self-trapped states of light carrying a
nontrivial angular momentum~\cite{discrete}. Similar localized
vortex states have been predicted to occur for BECs in optical
lattices~\cite{lena,tristram}.

Since many of {\em classical lattice systems} supporting discrete
vortices originate from {\em quantum models} or have a well-defined
quantum limit, the fundamental question is: {\em Do these coherent
structures survive under the action of quantum fluctuations?} In
this Letter we answer this question and reveal the intimate
connection between the classical discrete vortex (CDV) and its
quantum counterpart, which we call {\em quantum discrete vortex} or
{\em q-vortex)}  In some sense, these q-vortex states can be
compared with {\em quantum breathers} \cite{bishop,fleurov,dorignac}
which provide a quantum analog of the self-trapped states with {\em
localized energy} in discrete classical lattices.

In this Letter, we consider the simplest case of a three-site
Bose-Hubbard ring. First, reducing the model to the discrete
self-trapping equations~\cite{eilbeck} with a variational approach,
we find CDVs. Next, calculating the spatial correlations in the
ground states and the fidelity between the exact and variational
ground states, we study  the superfluid-insulator crossover and find
a valid regime of the classical variational approach. Last,
analyzing the phase coherence and quantum fluctuations of the
q-vortices, we find that the coherent structure ``melts" under the
action of quantum fluctuations, and this melting process accompanies
the superfluid-insulator crossover. This is in a sharp contrast to
CDVs, whose phase coherence is independent of the inter-site
coupling strength and the total number of particles in the vortex.

\begin{figure}[h]
\rotatebox{0}{\resizebox *{\columnwidth}{2.6cm} {\includegraphics
{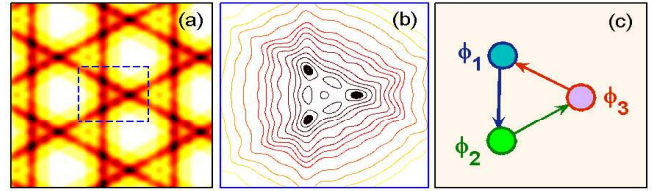}}} \caption{(colour online) Schematic diagrams. (a)
Bosons in a strongly trimerized Kagom$\acute{e}$ lattice
\cite{Kagome-Lattices} can be reduced to three-site Bose-Hubbard
rings. (b) Combining a proper two-dimensional harmonic potential
with the Kagom\'{e} lattice, a triple-well potential can be
generated; (c) three-site state.}
\end{figure}

The basic concepts of the quantum discrete vortices can be captured
by the simplest case of a three-site Bose-Hubbard ring, as shown
schematically in the diagrams of Fig.~1.  The triple-well potential
required for such a state may be readily found as a subset of the
familiar Kagom\'{e} lattices~\cite{Kagome-Lattices} [see
Fig.~1(a,b)]. For a deep potential, the three-site subsets become
decoupled from one another and each of them obeys the Hamiltonian,
\begin{equation}
\label{eq_1} H =-T \sum_{<l,m>} (a^{+}_la_m + a_l a^{+}_m) +
\frac{U}{2} \sum_{l=1}^3 n_l(n_l-1),
\end{equation}
where the operators $a^{+}_l$ and $a_l$ generate  and annihilate a
particle on the $l$-site, and $T$ and $U$ stand for the tunneling
and on-site interaction strengths, respectively. We consider a
repulsive inter-particle interaction ($U>0$) assuming the
application of our results to BECs.

Our first goal is to connect the classical and quantum pictures for
describing the stationary states of this system, and to this end we
use the time-dependent variational approach~\cite{TDVA-SU}. For the
three-site model, we introduce the SU(3) coherent state,
\begin{equation} \left| \Psi \right\rangle
=\frac{1}{\sqrt{N!N^{N}}}\left( \sum\limits_{l=1}^{3}\psi
_{l}a_{l}^{+}\right) ^{N}\left| vs\right\rangle,
\end{equation}
instead of a conventional product of Glauber's coherent states.
Here, $N$ is the total number of bosons, $\left| vs\right\rangle $
is the vacuum state $\left| n_{1}=0,n_{2}=0,n_{3}=0\right\rangle$,
and the complex amplitudes $\psi _{l}$ satisfy the normalization
condition, $\left| \psi _{1}\right| ^{2}+\left| \psi _{2}\right|
^{2}+\left| \psi _{3}\right| ^{2}=N$. In contrast to the variational
approach based on Glauber's coherent states \cite{TDVA}, this
variational approach conserves the total particle number $N$.
Minimizing the corresponding action, we derive the classical
Hamiltonian for the complex amplitudes $\psi_{l}$,
\begin{equation}
H_{\rm c}=-T\sum\limits_{<l,m>}(\psi _{l}^{*}\psi _{m}+\psi _{l}\psi
_{m}^{*})+ \frac{U_N}{2}  \sum\limits_{l=1}^{3}\left| \psi
_{l}\right| ^{4},
\end{equation}
where $U_N \equiv (N-1)U/N$.  The extra factor $(N-1)/N$ does not
appear in the Hamiltonian obtained from the variational approach
using Glauber's coherent states, and the complex amplitudes $\psi
_{l}$ obey the three-site discrete self-trapping equations
\cite{eilbeck,lederer,franzosi}:
\begin{equation}
i\frac{d\psi _{l}}{dt}=  - T \sum\limits_{m \neq l} \psi _{m} + U_N
|\psi _{l}|^{2} \psi _{l},
\end{equation}
with $\{l,m\}=1,2,3$. We are interested in the stationary
solutions and therefore make the substitution $\psi _{l} = A_{l}
\exp [-i(\mu t- \phi _{l})]$ with chemical potential $\mu$,
non-negative real amplitudes $A_{l}$ and phases $\phi_{l}$. The
system of equations (4) has a well-known degenerate set of
solutions with equal amplitudes~\cite{semiclassical-vortex}. The
solutions with $\phi_{1}=\phi_{2}=\phi _{3}$,
$A_{1}=A_{2}=A_{3}=\sqrt{N/3}$ and $\mu=(N-1)U/3-2T$ describe the
ground state of the system. The {\em symmetric vortex states} have
$A_{1}=A_{2}=A_{3}=\sqrt{N/3}$, $\phi_{2}-\phi_{1}=\phi_{3}-\phi
_{2}=2l\pi/3$ (where $l$ are nonzero integers) and
$\mu=(N-1)U/3-2T\cos(2l\pi/3)$. These three-site CDVs exist for
arbitrary values of $N$, $U$ and $T$. There exist also asymmetric
solutions which have only two equal amplitudes. These states are
the excited states of the system and ultimately connect with
localization maintained by self-trapping.

We now return to the fully quantum description for the Hamiltonian
(\ref{eq_1}), in which the behavior depends on the ratio $U/T$ and
$N$. The most prominent effect in infinite quantum systems is the
{\em superfluid-insulator transition}. In our finite-sized system,
the ground state is indeed superfluid in the strong tunneling limit,
$U/T \ll 1$, as expected from the infinite case. In the
weak-tunneling limit, $U/T \gg 1$, the ground state is insulating
for commensurate cases ($N=3k$ with a positive integer $k$) but has
a small superfluid fraction accompanying an insulating core for
incommensurate cases ($N \neq 3k$). However, instead of a sharp
phase transition observed for infinite systems, these two limits are
connected by {\em a crossover regime}. That is to say, the
superfluid-insulator crossover in this finite-site system with
commensurate fillings takes the place of the superfluid-insulator
transition predicted and observed in the infinite
systems~\cite{QPT}, and hence there is no well-defined critical
value of the ratio $U/T$ between the superfluid and insulating
phases. Using exact numerical diagonalization, we calculate the
spatial correlations $\left| \left\langle
a_{i}^{+}a_{i+1}\right\rangle \right| /n_{\rm av}$, where $n_{\rm
av}=N/3$ is the average number of particles per site, see Fig.~2.
These results show that in the limit of weak tunneling the
superfluid fraction  depends strongly on $N$ and the spatial
correlations decrease with $U/T$ such that the system approaches its
classical counterpart analyzed above.

\begin{figure}[h]
\rotatebox{0}{\resizebox *{\columnwidth}{6.0cm} {\includegraphics
{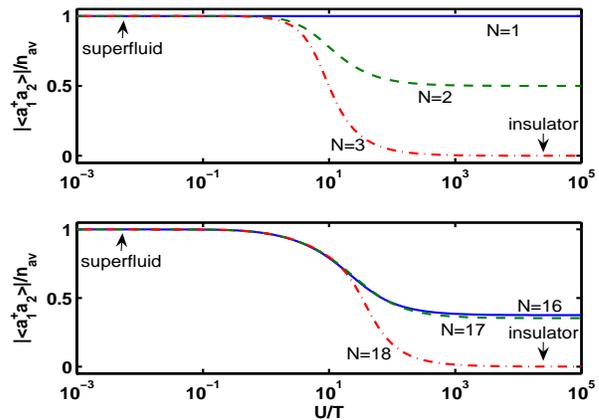}}} \caption{(colour online) Spatial correlations
vs.  $U/T$.}
\end{figure}

As the first step to compare the classical results found using the
variational approach with those obtained using exact
diagonalization, we compare the ground-state energies. Denoting
$E^{G}_{\rm va}$ the ground-state energy obtained from the
variational approach and $E^{G}_{\rm ex}$ that obtained by exact
diagonalization, we show that the fractional difference of the
ground state energies $(1-E^{G}_{\rm va}/E^{G}_{\rm ex})$ increases
with the ratio $U/T$ close to the linear limit, $U=0$. In the limit
of strong tunneling, i.e. for $U/T \ll 1$, the results show an
excellent agreement between the two approaches. This is also
confirmed by the fidelity between the exact and variational ground
states, $F=\left|_{\rm va}\left\langle
\text{GS}|\text{GS}\right\rangle _{\rm ex}\right| ^{2}$, which
decreases with $U/T$ for arbitrary $N$, see Fig.~3. With
$U/T<0.1995$, for $N$ from 1 to 18, $(1-E^{G}_{\rm va}/E^{G}_{\rm
ex})<8 \times 10^{-3}$ and $F>0.9790$. The fidelity decrease with
$U/T$ indicates that the breakdown of the variational approximation
accompanies the occurrence of the superfluid-insulator crossover.

\begin{figure}[h]
\rotatebox{0}{\resizebox *{\columnwidth}{6.0cm} {\includegraphics
{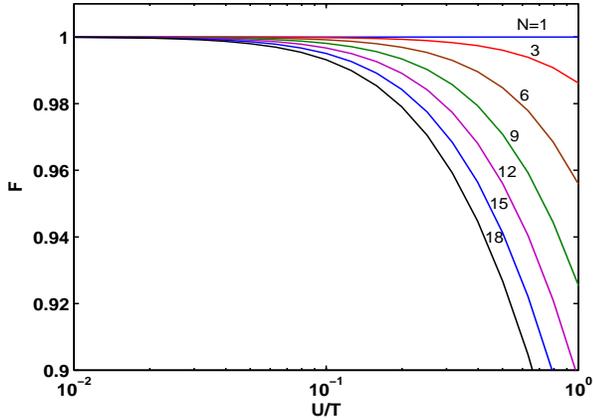}}} \caption{(colour online) Fidelities between the
variational and exact ground states vs. $U/T$ for different $N$.}
\end{figure}

Now we can construct {\em the quantum vortex states}. The fully
quantum counterparts for symmetric CDVs can be presented as SU(3)
coherent states $\left| \Psi \right\rangle$ with complex amplitudes
$\psi _{k}=\sqrt{N/3} \times \exp \left[ i\left( \varphi + 2(k-1)L
\pi /3 \right) \right]$ with $k (=1,2,3)$ and $L$ denote the site
index and the vortex charge, respectively. Such {\em q-vortices},
\begin{equation}
\left| \text{QDV} \right\rangle =
\sum\limits_{n_{2}=0}^{N}\sum\limits_{n_{3}=0}^{N-n_{2}}
\text{C}(n_{2},n_{3}) \left| n_{1},n_{2},n_{3}\right\rangle,
\end{equation}
can be prepared in the limit of strong tunneling by imprinting the
specific phase structure $(\varphi, \varphi + 2L\pi /3, \varphi +
4L\pi /3)$ onto the ground states, $ \left| \text{GS} \right \rangle
=\sum \sum \text{G}(n_{2},n_{3}) \left|
n_{1},n_{2},n_{3}\right\rangle $, and the coefficients satisfy
\[
\text{C}(n_{2},n_{3})=\text{G}(n_{2},n_{3}) \exp \left\{i[N
\varphi + \frac{2}{3}n_{2}L\pi + \frac{4}{3}n_{3}L\pi]\right\},
\]
where $N=n_{1}+n_{2}+n_{3}$. Fast phase imprinting for an atomic
cloud with a required phase structure via atom-laser interaction was
also suggested for entanglement preparation
\cite{Reinhardt-Entanglement}, as well as atomic dark-soliton
generation and measurement~\cite{phase-imprint}. The asymmetric
q-vortices can be obtained by applying similar phase imprinting
procedures to excited states whose classical limit (strong tunneling
limit) correspond to asymmetric solutions. Below, we consider only
symmetric vortices.

In the linear limit, the q-vortices with $L=3k$ ($k$ are integers)
are ground states of the system, while those with $L=3k \pm 1$
correspond to the excited states with higher energies. Beyond the
linear limit, the quantum counterparts for CDVs can be obtained by
the quantum adiabatic evolution. For our small-size systems we
simulate the time evolution in the complete Hilbert space with the
Runge-Kutta integration scheme~\cite{evolution}.  We change the
parameters adiabatically such that the populated state is always
very close to an eigenstate.  To determine whether a given state
or even the quantum vortex itself is a well-defined eigenstate, we
project it onto a complete Hilbert basis of an ensemble of the
orthogonal eigenstates. In the linear limit, we find that all
nonzero-probability components of the vortex {\em have identical
eigenvalues} and, therefore, we may conclude that the q-vortex is
an eigenstate. Adiabatically varying $U/T$ to the strongly
nonlinear limit, these q-vortices evolve into different final
states dependent on the charges $L$. For $L=3k$, the final states
are single-peaked states in the distribution of the probability
amplitudes $|\text{C}(n_{1},n_{2})|^{2}$ (the ground state).
However, for $L = 3k \pm 1$, the quantum vortex (5) appears to be
a well-defined eigenstate with a desired probability $P_{es}$ if
$U/T$ is less than a certain value. For $N=6$, $P_{es}>0.9850$
when $U/T < 0.1259$. As the nonlinearity is adiabatically
increasing, the q-vortex breaks down ending up in the limit of
strong nonlinearity as a triple-peaked state in the probability
distribution. The appearance of the single-peaked and
triple-peaked states is independent of total atom number, it
indicates the loss of the circular current of particles and
effective melting of the discrete vortex structure. In Fig.~4, we
show the quantum adiabatic evolution of the q-vortices with $N=6$.

\begin{figure}[h]
\rotatebox{0}{\resizebox *{\columnwidth}{6.0cm}{\includegraphics
{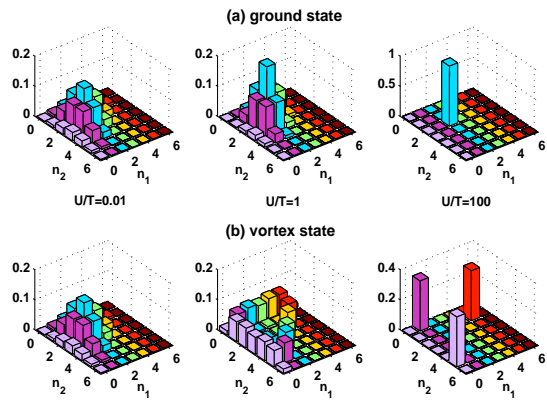}}} \caption{(colour online) Quantum adiabatic evolution
of q-vortices with $N=6$ from linear to nonlinear limit. (a) and (b)
correspond to the cases of $L=3k$ and $L=3k \pm 1$, respectively.}
\end{figure}

To explore in detail how the classical discrete vortices ``melt", we
calculate {\em quantum fluctuations} of their quantum counterparts,
i.e., the q-vortices. Using the quantum phase concept
\cite{q-phase}, we introduce the cosine and sine functions of the
quantum phase $\phi _{j}$ for the $j-$th site as $\cos \phi _{j} =
\frac{K_{j}}{2}(a_{j}^{+}+a_{j})$ and $\sin \phi _{j} =
\frac{iK_{j}}{2}(a_{j}^{+}-a_{j})$ with the constant $K_{j}$
determined by the particle number. Thus, the cosine and sine
functions for the two-body phase difference $\phi = \phi _{2}-\phi
_{1}$ can be defined as,
\begin{equation}
\cos \, \phi = K (a_{2}^{+}a_{1} + a_{2}a_{1}^{+}), \sin \, \phi = i
K (a_{2}^{+}a_{1} - a_{2}a_{1}^{+}),
\end{equation}
where the constant $K=K_{1}K_{2}$. Using the conservation character
of the square summary, $\left \langle \sin^{2}\phi + \cos^{2}\phi
\right\rangle=1$, it is easy to obtain $K_{1}K_{2} =
1/\sqrt{2\left\langle 2n_{1}n_{2}+n_{1} + n_{2} \right \rangle}$.

The expectation value and the variance of $\cos \phi$ are
\begin{equation}
\left\langle \cos \, \phi \right\rangle =\frac{\left\langle
a_{2}^{+}a_{1}+a_{2}a_{1}^{+}\right\rangle }{\sqrt{2\left\langle
2n_{1}n_{2}+n_{1}+n_{2}\right\rangle }},
\end{equation}
and
\begin{equation}
\Delta (\cos \, \phi) = \left\langle \cos ^{2}\phi\right\rangle
-\left\langle \cos \phi\right\rangle ^{2},
\end{equation}
respectively. Here, the expectation value for $\cos^\phi$ is
\begin{equation}
\left\langle \cos ^{2}\phi \right\rangle =\frac{1}{2} +
\frac{\left\langle (a_{2}^{+}a_{1})^{2}\right\rangle +\left\langle
(a_{2}a_{1}^{+})^{2}\right\rangle }{2\left\langle
2n_{1}n_{2}+n_{1}+n_{2}\right\rangle }.
\end{equation}
If the matter waves located at two neighboring sites have a
well-defined phase difference $\phi$, the variance of $\cos \phi$
vanishes. Otherwise (e.g., the values of $\phi$ are random), the
expectation value for $\cos \phi$ and the corresponding variance
will approach zero and 1/2, respectively. The random distribution of
the phase difference means `melting' of the coherence between the
matter-wave clouds localized at the neighboring sites.

\begin{figure}[h]
\rotatebox{0}{\resizebox *{\columnwidth}{6.0cm} {\includegraphics
{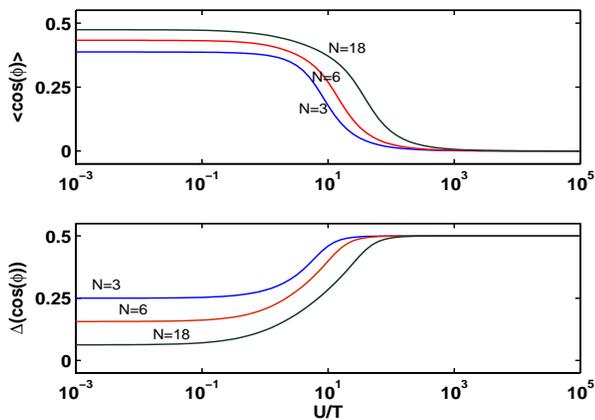}}} \caption{(colour online) Quantum fluctuations for
the vortex states vs. the ratio $U/T$ for different values of $N$.}
\end{figure}

For the q-vortices (5) with the charge $L=1$, by adiabatically
increasing $U/T$ we  calculate numerically the expectation values
$\left\langle \cos \phi \right\rangle$ and variances $\Delta (\cos
\phi)$ between the sites 1 and 2 (where $\cos \phi = \cos (\phi
_{2}-\phi _{1})$) for different $U/T$ starting from the linear
limit. For the commensurate cases, the expectation value
$\left\langle \cos \phi \right\rangle$ grows with $N$ and decreases
with $U/T$, but the variance $\Delta (\cos \phi)$ decreases with $N$
and increases with $U/T$, as shown in Fig.~5. These results mean
that quantum fluctuations decrease with $N$ but increase with $U/T$.
In the limit of strong tunneling, i.e. for $U/T \ll 1$, the
expectation values $\left\langle \cos \phi \right\rangle \rightarrow
\cos (2\pi/3)$ and variances $\Delta (\cos \phi) \rightarrow 0$ for
$N \rightarrow \infty$. The limit of {\em strong tunneling} with
large total particle numbers corresponds to {\em the classical
limit} with well-defined phases. The opposite occurs in the limit of
weak tunneling ($U/T \gg 1$) where the expectation values
$\left\langle \cos \phi \right\rangle \rightarrow 0$ and variances
$\Delta (\cos \phi) \rightarrow 1/2$ for arbitrary $N$. This means
that the distribution of the phase differences becomes random, and
the classical discrete vortices melt completely under the action of
strong quantum fluctuations. Varying $U/T \ll 1$ to $U/T \gg 1$, we
recover the crossover regime that connects these two limits.

We suggest that these effects can be studied experimentally by
loading an $^{87}\text{Rb}$ condensate into the triple-well
potential formed by a superposition of a harmonic potential (with
frequency $\omega \sim 300 \times 2\pi \text{Hz}$) and a Kagom\'{e}
lattice~\cite{Kagome-Lattices}, see Fig. 1. The Kagom\'{e} lattice
can be formed by laser beams with wavelengths of 1064 nm. The ratio
$T/U$ can then be adjusted by varying the laser intensity about from
1.5 to 15 $\text{W/cm}^{2}$ to observe the crossover regime and the
melting of the vortex phase.

In conclusion, we have studied a quantum analog of discrete vortices
as the states of interacting bosons with a nontrivial phase
structure. We have introduced the q-vortices for the simplest case
of a three-site Bose-Hubbard ring and analyzed the effect of quantum
fluctuations on these states. We have found that the melting of
discrete vortices via quantum fluctuations accompanies the crossover
from superfluid to Mott insulator. We believe our findings may
initiate experimental efforts to observe quantum discrete vortices
in Bose-Einstein condensates in optical lattices.

We thank E. A. Ostrovskaya,  A. A. Sukhorukov, L. Santos, Y. -Z.
Zhang, P. Drummond, and J. Corney for discussions, R. Gati for
experimental parameters estimation, and the Australian Research
Council for support.

\end{document}